\documentclass[aps,prb,reprint,superscriptaddress]{revtex4-1}
\usepackage{amsmath}
\usepackage{graphicx}
\usepackage{natbib}

\begin{document}

\title{Nonlinear Transient Dynamics of Photoexcited Silicon Nanoantenna for Ultrafast All-Optical Signal Processing}

\author{Denis G. Baranov}
\email[]{denis.baranov@phystech.edu}
\affiliation{Moscow Institute of Physics and Technology, 9 Institutskiy per., Dolgoprudny 141700, Russia}

\author{Sergey V. Makarov}
\affiliation{ITMO University, Saint-Petersburg, Russia}

\author{Valentin A. Milichko}
\affiliation{ITMO University, Saint-Petersburg, Russia}

\author{Sergey I. Kudryashov}
\affiliation{ITMO University, Saint-Petersburg, Russia}

\author{Alexander E. Krasnok}
\affiliation{ITMO University, Saint-Petersburg, Russia}

\author{Pavel A. Belov}
\affiliation{ITMO University, Saint-Petersburg, Russia}

\date{\today}


\begin{abstract}
Optically generated electron-hole plasma in high-index dielectric nanostructures was demonstrated as a means of tuning of their optical properties. However, until now an ultrafast operation regime of such plasma driven nanostructures has not been attained. Here, we perform pump-probe experiments with resonant silicon nanoparticles and report on dense optical plasma generation near the magnetic dipole resonance with ultrafast (about 2.5 ps) relaxation rate. Basing on experimental results, we develop an analytical model describing transient response of a nanocrystalline silicon nanoparticle to an intense laser pulse and show theoretically that plasma induced optical nonlinearity leads to ultrafast reconfiguration of the scattering power pattern. We demonstrate 100 fs switching to unidirectional scattering regime upon irradiation of the nanoparticle by an intense femtosecond pulse. Our work lays the foundation for developing ultracompact and ultrafast all-optical signal processing devices.
\end{abstract}

\maketitle

\section{Introduction}

Silicon presents a versatile and low-cost platform for data processing at speeds of 100~Gbit/s and beyond.\cite{Leuthold} This is possible thanks to a broad range of inherent optical nonlinearities such as stimulated Raman scattering,\cite{Rong2005} Kerr effect,\cite{Koos} two-photon absorption,\cite{Tanabe} thermo-optical effect,\cite{Notomi} and electron-hole plasma (EHP) generation.\cite{Almeida} 
However, the relatively large size of such devices and their high power consumption have remained an obstacle.


Miniaturization of an optical device is possible with use of high-index dielectric (including silicon) nanoparticles which due to their Mie resonances enhance light-matter interaction at the nanoscale.\cite{popa2008compact, Geffrin2012, rolly2012boosting, Kuznetsov2012, Evlyukhin2012, krasnok2012all, Lukyanchuk2015} 
Recently, such all-dielectric nanostructures have been studied in the context of huge enhancement of optical nonlinearities.\cite{Shcherbakov2014,Lewi2015, Makarov2015,Shcherbakov2015, yang2015nonlinear} The ability to control the scattering behavior of nanoparticles via nonlinear effects may open unique opportunities for effective light manipulation by using only a single dielectric nanoparticle.

\begin{figure}[!b]
\includegraphics[width=1\columnwidth]{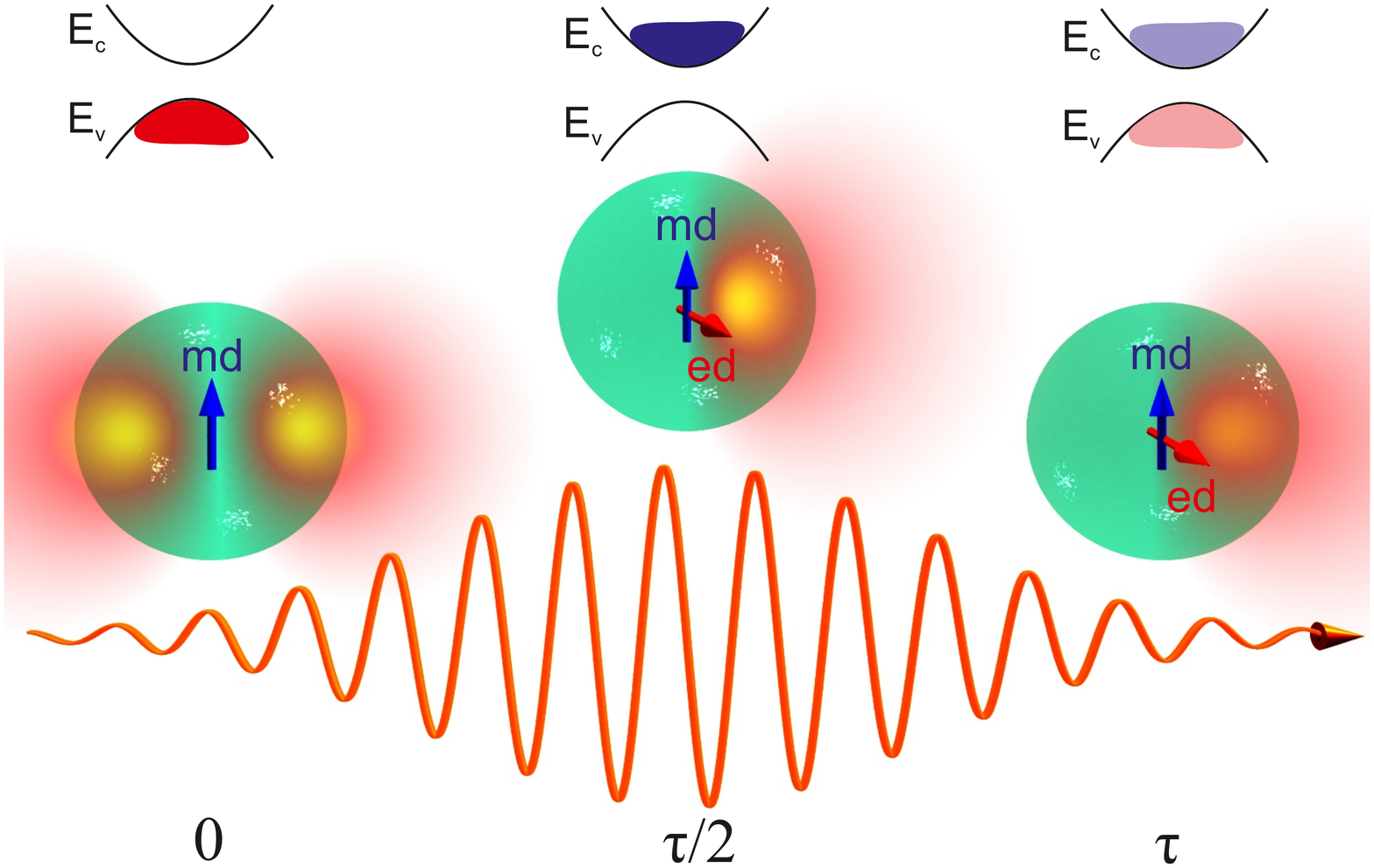}
\caption{Schematic illustration of transient dynamics of a silicon nanoparticle scattering properties during its photoexcitation by a femtosecond laser pulse with duration $\tau$. Optical absorption causes electrons to fill the conduction band of silicon, therefore modifying its permittivity and nanoantenna radiation pattern.}
\label{fig1}
\end{figure}

The choice of optical nonlinearity for data processing is an important factor, which determines the ultimate performance of a photonic device. Refractive index change induced by EHP nonlinearity can be significantly larger than that of Kerr nonlinearity,\cite{Leuthold} avoiding the necessity to use high-$Q$ large resonators and photonic crystals. The generation of EHP was employed for tuning of silicon nanoantenna optical properties in the IR and visible regions.\cite{Lewi2015,Makarov2015} However, in contrast to the instantaneous Kerr mechanism, EHP relaxation process at low-intensity photoexcitation is relatively slow (tens-hundreds of picoseconds), being comparable with the fastest electronic devices. Such slow EHP relaxation in silicon-based nanoantennas was demonstrated in previous studies,\cite{Shcherbakov2015, yang2015nonlinear} where ultrafast (sub-picosecond) reconfiguration of resonant silicon (Si) nanoparticles was achieved only via relatively weak Kerr-type optical nonlinearity. Therefore, for achieving effective all-optical signal processing by means of a single dielectric nanoparticle, two basic problems should be solved: (i) achieving of a few picosecond time of EHP relaxation, and (ii) optimization of the modulation depth and time of switching.


\begin{figure}[!t]
\includegraphics[width=0.9\columnwidth]{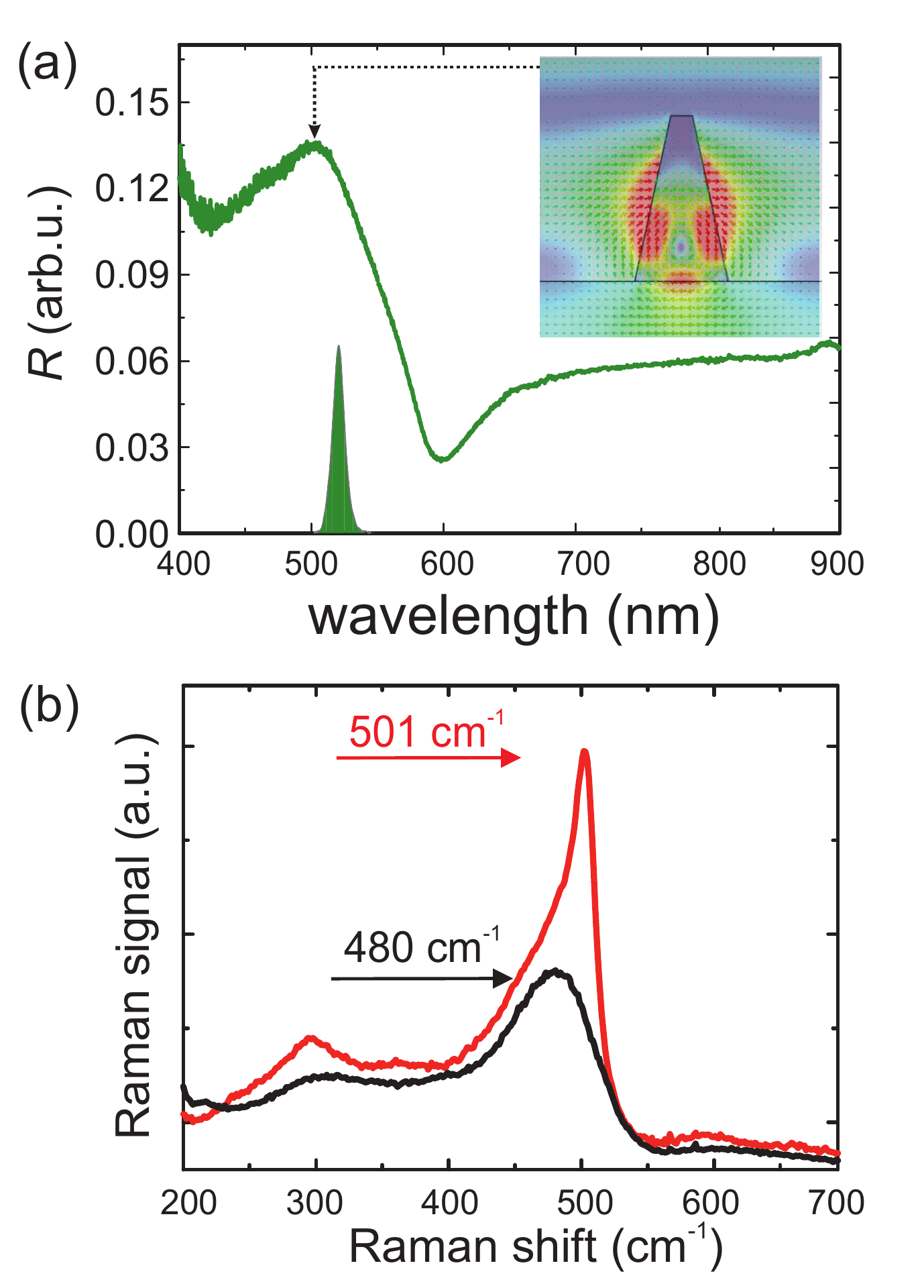}
\caption{(a) Reflection spectrum of cone silicon nanoparticles with height 220~nm, and bottom/top diameters of 140~nm/40~nm. (b) Raman spectra of initial Si:H film (black curve) and nanoparticles fabricated from the film (red curve).}
\label{fig2}
\end{figure}


In this Paper, we report on experimental observation of $\sim$2.5~ps operation regime of a nonlinear all-dielectric nanoantenna, which is one order of magnitude faster than previously reported for resonant Si nanoparticles. Basing on the pump-probe experiments, we build an analytical model describing the transient dynamics of optical scattering on a dielectric nanoparticle with magnetic and electric optical responses. The developed model allows us to analyze the excitation and radiation of the electric and magnetic dipole modes in strongly nonlinear regime of EHP generation and ultrafast relaxation. In particular, we describe reconfiguration of radiation pattern during the pulse action from dipole-like to unidirectional regime mediated by generation of EHP in the nanoparticle, schematically shown in Fig.~\ref{fig1}.

\section{Pump-probe measurements}
In order to demonstrate the regime of ultrafast silicon nanoantenna operation, we performed a series of pump-probe experiments with an array of silicon nanoparticles supporting magnetic optical response. The cylindrical nanoparticles have bottom/top diameters of 140~nm/40~nm, a height of 220~nm, and a period of 600~nm.
Results of linear optical measurements [see Fig.~\ref{fig2}(a)] show pronounced resonance near 500~nm which is attributed to magnetic dipole Mie resonance of a single nanoparticle. The resonance response is confirmed by our numerical modeling in CST Microwave Studio, revealing typical ring-like electric field distribution within the nanoparticle [see inset in Fig.~\ref{fig2}(a)].

The nanoparticles were fabricated by means of plasma-enhanced chemical vapour deposition in SH$_{3}$ atmosphere, which usually results in the formation of completely amorphous Si:H material with typical broad maximum at 480~cm$^{-1}$ in Raman spectrum [Fig.~\ref{fig2}(b), black curve]. However, at the final stage of the lithographical procedure the resulting nanoparticles represent partially recrystallized state with the narrow Raman peak at 501~cm$^{-1}$ with asymmetric shape caused by influence of the remain amorphous component~\cite{he1994structure} [Fig.~\ref{fig2}(b), red curve]. According to our Raman micro-spectroscopy measurements (for details see Supplementary materials), the amount of crystalline fracture $I_{c}/(I_{c}+I_{a})$ is about 0.45, where $I_{c}$ and $I_{a}$ are the integrated intensities of the crystalline and amorphous phases. The partial recrystallization of initially amorphous Si:H film is supposed to be caused  by electron-beam processing step with 25~kV acceleration voltage.\cite{jencic1995electron}
The resulting pertially recrystalized material has average grain size  about d~$\approx$~2 nm,
 according to the expression ${\text{d}} = 2\pi \sqrt {{\text{B}}/\Delta \omega } $, where $\Delta\omega$ is the peak shift for the microcrystalline as compared that of the c-Si, and 
B~=~2~cm$^{-2}$.\cite{he1994structure} Such material properties are preferable for enhancement of radiative recombination of generated EHP carriers, reducing thermal recombination.\cite{zacharias2002size}

\begin{figure}[!t]
\includegraphics[width=0.83\columnwidth]{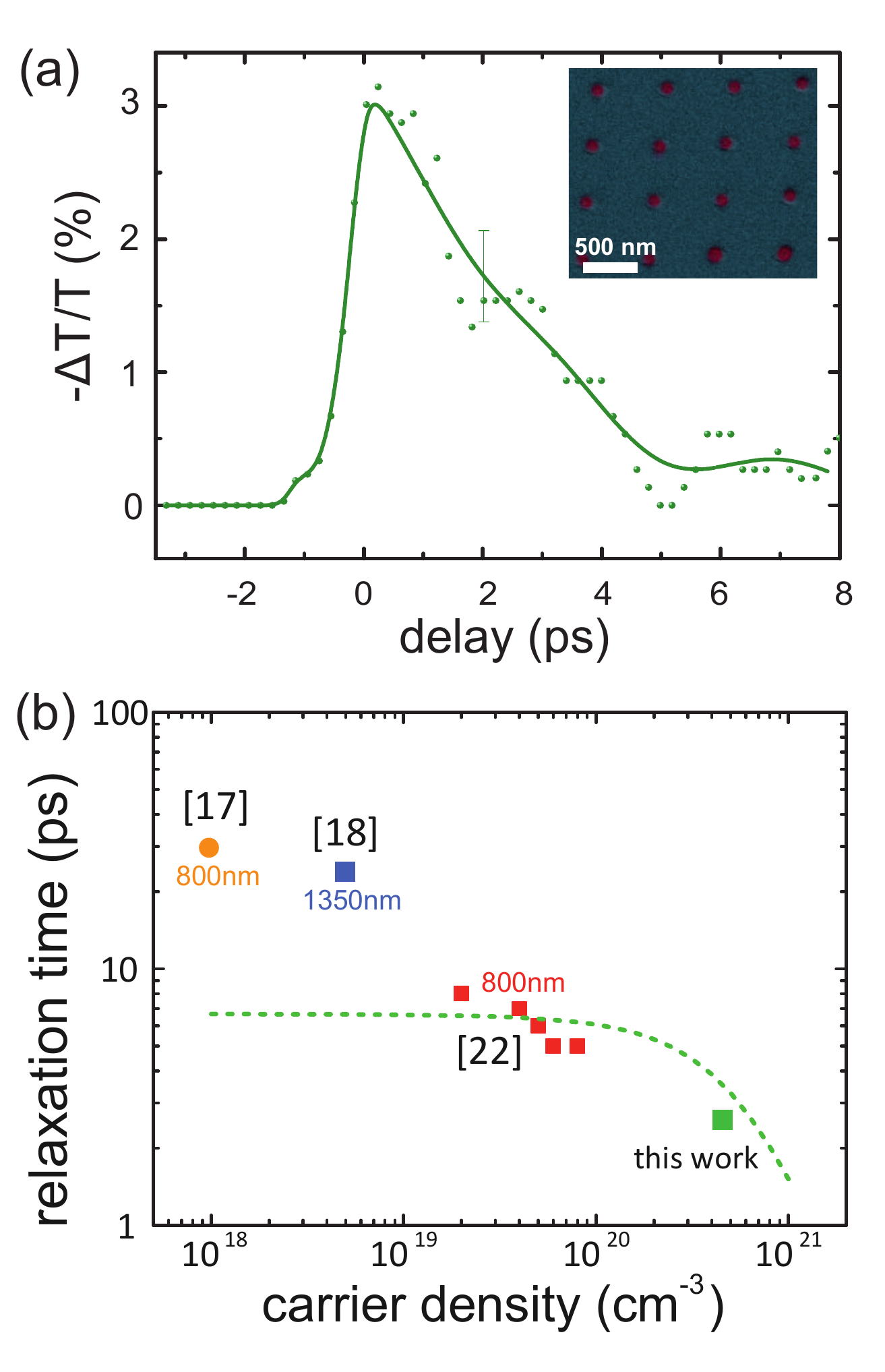}
\caption{(a) Relative transmittance change as a function of probe delay. Inset: colorized SEM image of the array of Si nanoparticles. (b) Electron-hole plasma relaxation times for resonant nanoparticles made of a-Si:H at pump wavelength 800~nm (orange circle~\cite{Shcherbakov2015}), for resonant nanoparticles made of pc-Si at 1350 nm (blue square~\cite{yang2015nonlinear}), for nonresonant nanocrystals inside a-Si:H bulk matrix at 800~nm (red squares~\cite{Optlett15}), and relaxation time measured in this work at 515~nm for resonant nanoparticles made of nc-Si/a-Si:H~(green square). Green dotted line corresponds to our model for the nc-Si/a-Si:H material.}
\label{fig3}
\end{figure}

Time-domain measurements of the transmission changes at high laser intensity were performed for testing the nonlinear properties of the nanoparticles. The experimental results are presented in Fig.~\ref{fig3}(a), where the relative transmittance change ${{\Delta T} \mathord{\left/ {\vphantom {{\Delta T} {{T_0}}}} \right. \kern-\nulldelimiterspace} {{T_0}}}$ of the probe beam through the array is shown as a function of the delay between two pulses $\Delta \tau$ with respect to the pump beam. Pulse duration of the pump beam is 220~fs, intensity is 200$\pm$100~GW/cm$^{2}$, and repetition rate is 1~kHz. This measurement at the highest possible intensity in the non-damage regime provides us the information on EHP relaxation time. Fitting the transmission transient in Fig.~\ref{fig3}(a) results preferably in decay time of $\approx 2.5 \pm 0.5$~ps. Relatively small values of modulation depth is described by slight mismatching of pump and probe beam on 15$\times$15~$\mu$m$^2$ nanoparticle array. We did not observe sharp two-photon absorption peak reported previously,\cite{Shcherbakov2015,yang2015nonlinear} probably, due to crossed polarizations of the pump and probe pulses excluding any interference effects. Nevertheless, the observed value of 2.5~ps is the smallest observed for EHP decay in silicon-based resonant nanoparticles, compared to $\approx$30~ps~\cite{Shcherbakov2015} and $\approx$25~ps~\cite{yang2015nonlinear} [Fig.~\ref{fig3}(b)]. The comparison of the results for the nanoparticles with those for bulk amorphous and nanocrystalline silicon phases is given in Supplementary materials.

\section{Analytical theory}

Now we derive the master equations describing temporal response of a Si nanoparticle exposed to a short optical pulse.
We model a spherical nanoparticle as a combination of electric ($\mathbf{p}$) and magnetic ($\mathbf{m}$) dipoles which are related to incident monochomatic electromagnetic fields via ${\mathbf{p}} = {\alpha _e}{\mathbf{E}}$ and ${\mathbf{m}} = {\alpha _m}{\mathbf{H}}$. The corresponding dipole polarizabilities are expressed in terms of the Mie coefficients $a_1$ and $b_1$ as ${\alpha _e} = \frac{{3i}}{{2{k^3}}}{a_1}$ and ${\alpha _m} = \frac{{3i}}{{2{k^3}}}{b_1}$, respectively,\cite{Bohren} where $k=\omega/c$ is the free space wave vector. The above expressions relate incident field and induced dipole moments in the stationary regime.

In order to obtain dynamical equations for dipole moments it is convenient to rewrite these expressions as $\alpha _e^{ - 1}{\mathbf{p}} = {\mathbf{E}}$ and $\alpha _m^{ - 1}{\mathbf{m}} = {\mathbf{H}}$.
Assuming that the spectrum of incident pulse is centered at the frequency $\omega_0$, we further represent incident waves as
${\mathbf{E}}\left( t \right) = \mathbf{\tilde E}\left( t \right)\exp \left( { - i{\omega _0}t} \right)$ and
 ${\mathbf{H}}\left( t \right) = \mathbf{\tilde H}\left( t \right)\exp \left( { - i{\omega _0}t} \right),$
where $\mathbf{\tilde E }\left( t \right)$ and $\mathbf{\tilde H } \left( t \right)$ denote slowly varying amplitudes of electromagnetic field.
Correspondingly, nanoparticle dipole moments read ${\mathbf{p}}\left( t \right) = {\mathbf{\tilde p}}\left( t \right)\exp \left( { - i{\omega _0}t} \right)$
and ${\mathbf{m}}\left( t \right) = {\mathbf{\tilde m}}\left( t \right)\exp \left( { - i{\omega _0}t} \right)$.
 Expanding inverse polarizabilities into Taylor series $\alpha _{e,m}^{ - 1} = \alpha _{e,m}^{ - 1}\left( {{\omega _0}} \right) + \frac{{\partial \alpha _{e,m}^{ - 1}}}{{\partial \omega }}\left( {\omega  - {\omega _0}} \right)$,
and performing Fourier transform of expressions relating induced dipole moments to the incident fields, we obtain the differential equations which dictate the optical dynamics:
\begin{equation}
\begin{gathered}
  i\frac{{\partial \alpha _E^{ - 1}}}{{\partial \omega }}\frac{{d{\mathbf{\tilde p}}}}{{dt}} + \alpha _E^{ - 1}{\mathbf{\tilde p}} = {\mathbf{\tilde E}}\left( t \right), \hfill \\
  i\frac{{\partial \alpha _H^{ - 1}}}{{\partial \omega }}\frac{{d{\mathbf{\tilde m}}}}{{dt}} + \alpha _H^{ - 1}{\mathbf{\tilde m}} = {\mathbf{\tilde H}}\left( t \right) \hfill \\ 
\end{gathered}
\label{eq1}
\end{equation}
The above equations fully describe dynamics of the electric and magnetic dipole moments provided that EHP density, which determines dipole polarizabilities, is known at all times. However, in our case the field intensity itself describes EHP generation, which we address below.

In order to describe dynamics of laser induced EHP, we employ the rate equation which can be found in literature.\cite{Zhang2015,Sokolowski-Tinten2000,Rong2004} For the sake of simplicity, we will describe volume averaged concentration $\rho_{\rm eh}(t)$ neglecting its spatial distribution and therefore diffusion of carriers across the particle volume. Indeed, at $\rho_{\rm eh}>10^{20}~{\rm cm}^{\rm -3}$ thermal velocity of hot free electrons is about $v_{e}\approx~5~\times~10^{5}$~cm/s,\cite{derrien2013possible} whereas corresponding electron-electron scattering time is about $\sim$100~fs. Tacking into account, the ring-like magnetic dipole mode structure, the characteristic EHP homogenisation time governed by a ballistic motion of electrons can be estimated as ${\tau _{\rm hom}} \sim R/2v_e \sim 100$~fs, where $R$ is the nanoparticles radius. We also ignore thermal effects, because of their negligible influence on epsilon at ps-timescale under non-damage irradiation conditions.

The rate equation for EHP therefore takes the form
\begin{equation}
\frac{{d{\rho _{{\text{eh}}}}}}{{dt}} =  - \Gamma {\rho _{{\text{eh}}}} + \frac{{{W_1}}}
{{\hbar \omega }} + \frac{{{W_2}}}{{2\hbar \omega }}.
\label{eq3}
\end{equation}
Here, $W_{1,2}$ are the volume-averaged dissipation rates due to one- and two-photon absorption and $\Gamma$ is the phenomenological EHP relaxation rate constant. The absorption rates are written in the usual form as ${W_1} = \frac{\omega }{{8\pi }}  \left\langle {{{\left| {{\mathbf{\tilde E_{\rm in}}}} \right|}^2}} \right\rangle {\rm Im} (\varepsilon)$
 and ${W_2} = \frac{\omega }{{8\pi }} \left\langle {{{\left| {{\mathbf{\tilde E_{\rm in}}}} \right|}^4}} \right\rangle \operatorname{Im} {\chi ^{(3)}}$, where $\left\langle {...} \right\rangle $ denotes averaging over the nanoparticle volume.
Nonlinear susceptibility $\operatorname{Im} {\chi ^{(3)}}$ may be expressed through experimentally accessible two-photon absorption coefficient $\beta$ via $\operatorname{Im} {\chi ^{(3)}} = \frac{{\varepsilon {c^2}}}{{8\pi \omega }}\beta $, where $\beta$ is a function of a pump laser wavelength.\cite{Reitze} These averaged fields should be related to the instantaneous values of electric ($\mathbf{\tilde p}$) and magnetic ($\mathbf{\tilde m}$) dipole moments. This is done by integrating the total field of the two spherical harmonics corresponding to the given values of $\mathbf{\tilde p}$ and $\mathbf{\tilde m}$.

\begin{figure*}[!t]
\includegraphics[width=2\columnwidth]{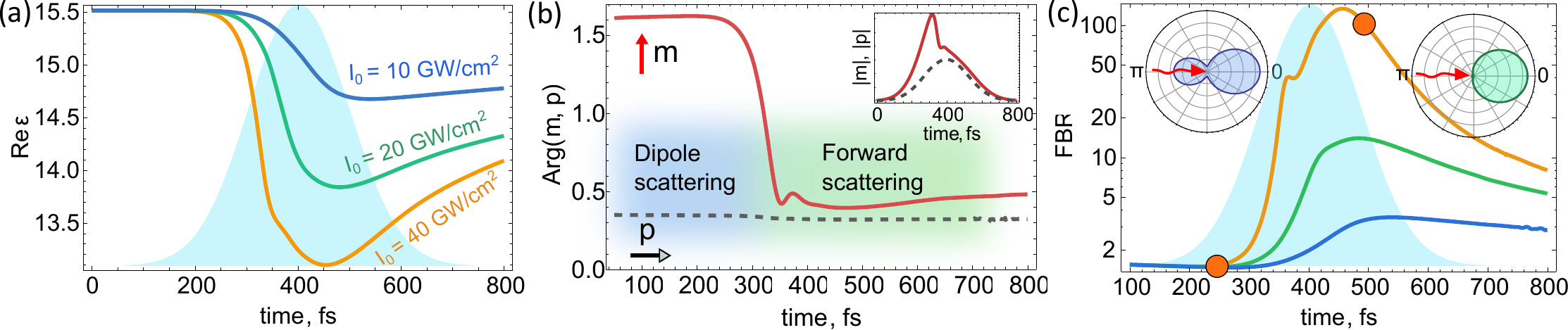}
\caption{Optical dynamics of a Si nanoparticle upon EHP generation. (a) Temporal dynamics of photoexcited Si permittivity for a particle irradiated by a resonant Gaussian pulse of different intensities. Shaded area represents the envelope of the pulse intensity. (b) Dynamics of ED and MD phasors excited in the particle during the pulse action for largest pulse intensity. Inset: the corresponding amplitudes of ED and MD. (c) Dynamical reconfiguration of nanoantenna directivity: FBR of a nanoparticle during the pulse action. Scattering diagrams of the incident beam at largest intensity are shown in the two insets.}
\label{fig4}
\end{figure*}

Generally, the total EHP relaxation rate ($\Gamma$) can be represented as polynomial function of EHP density: $\Gamma~=~{\Gamma _{{\text{TR}}}(\rho_{\rm eh}^{0})} + {\Gamma _{\text{BM}}(\rho_{\rm eh}^{1})} + {\Gamma _{\text{A}}(\rho_{\rm eh}^{2})}$, where each term corresponds to trapping ($\Gamma _{\rm TR}$), bi-molecular ($\Gamma _{\rm BM}$), and Auger ($\Gamma _{\rm A}$) mechanisms. Our experimental results allow us to take realistic parameters for our model. We describe the observed ultrafast recombination basing on the measurements of nc-Si from Ref.~\cite{Optlett15} with including of Auger recombination mechanism, because of mixed a-Si/nc-Si composition of our material.\cite{Shank} Therefore the resulting relaxation rate is represented via following terms: $\Gamma_{\rm TR}$~=~1.5~$\cdot$~10$^{11}$~s$^{-1}$ (Ref.~\cite{Optlett15}), $\Gamma_{\rm BM}$~=~1.1~$\cdot$~10$^{-10}$~$\rho_{\rm eh}$~cm$^3$s$^{-1}$ (Ref.~\cite{Optlett15}) and $\Gamma_{\rm A}$~=~4~$\cdot$~10$^{ - 31}$~$\rho_{\rm eh}^2$~s$^{-1}$ (Ref.~\cite{Shank}). The resulting dependence of relaxation time on EHP density is shown in Fig.~\ref{fig3}(c), demonstrating good agreement with the experimental values.

The last essential element of our dynamical model is the expression relating permittivity of excited silicon $\varepsilon$ to EHP density $\rho_{\rm eh}$. Generally, this dependence is represented as a following expression:\cite{Sokolowski-Tinten2000, Makarov2015}
\begin{eqnarray}
{\varepsilon }\left( {\omega ,{\rho _{{\text{eh}}}}} \right) = \varepsilon_{\text{Si}}+ \Delta\varepsilon_{\rm bgr}+ {\Delta\varepsilon _{{\text{bf}}}} + {\Delta\varepsilon _{{\text{D}}}}
\label{eq6}
\end{eqnarray}
where ${\varepsilon _{{\text{Si}}}}$ is the permittivity of non-excited material, whereas $\Delta\varepsilon_{\text{bgr}}$, ${\Delta\varepsilon _{{\text{bf}}}}$, and $\Delta\varepsilon _{{\text{D}}}$ are the contributions from band gap renormalization, band filling, and Drude term. The Drude contribution strongly dominates the others in the IR range and at relatively low intensities.\cite{Leuthold} Band filling is important when EHP density becomes comparable with the capacity of the conduction band ($\rho_{\rm eh}~>$~10$^{20}$~cm$^{-3}$). Band gap renormalization plays significant role at wavelengths where permittivity dispersion d$\varepsilon$/d$\omega$ is considerable ($<$~800~nm for Si). The detailed expressions for all contributions in Eq.~(\ref{eq6}) are given in Supplementary materials.

In the simulations, ${\varepsilon(\rho_{\rm eh})}$ becomes time-dependent and determines inverse dipole polarizabilities entering electromagnetic part of our model, Eq.~(\ref{eq1}). The system of equations (\ref{eq1}) and (\ref{eq3}) allow to completely determine transient behavior of a silicon nanoparticle under action of an intense optical pulse of an arbitrary shape.


\begin{figure*}[!t]
\includegraphics[width=1.9\columnwidth]{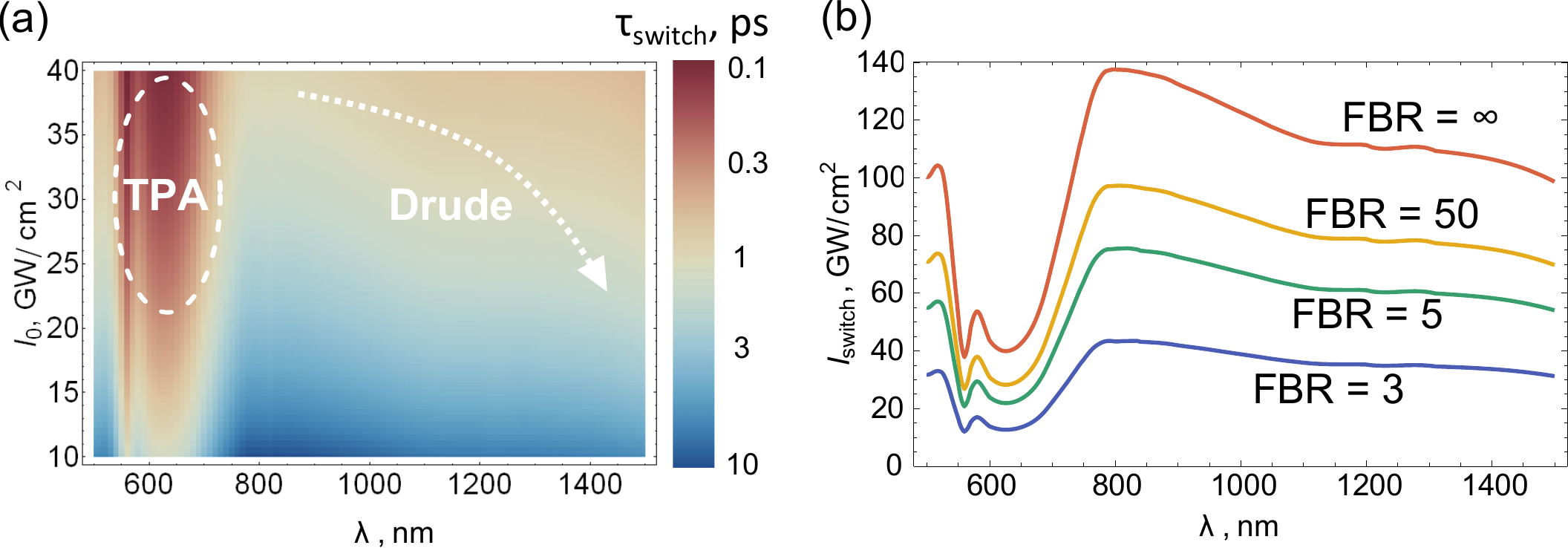}
\caption{(a) Map of switching time $\tau_{\rm switch}$ as a function of incident pulse wavelength and peak intensity. (b) Wavelength dependence of the switching intensity (for 200~fs pulse) calculated for a series of FBR values.}
\label{fig5}
\end{figure*}

\section{Results}
We now apply our model to study the interplay of excited electric and magnetic dipoles, leading to directional scattering.\cite{Kerker} We illustrate the dynamical reconfiguration of the scattering pattern for the R~=~75~nm radius silicon particle in air. The particle is driven by a Gaussian pulse at wavelength $\lambda=600$ nm which is tuned to magnetic dipole resonance, where $\beta$ of silicon has the highest values,\cite{Reitze} allowing to reduce incident pulse intensity. In Fig.~\ref{fig4} we present detailed transient optical response for a few peak incident intensities $I_0$ determined as $I_0=(c/8\pi)|E|^2$. Pulse duration is taken as $\tau=200$~fs, pulse center position is $t_0=400$ fs.

Figure~\ref{fig4}(a) shows the time-dependent real part of Si permittivity for various $I_0$. It shows a pronounced dip near the pulse center around $t=500$~ps, where EHP density is maximal.
Electric dipole (ED) and magnetic dipole (MD) phases for the largest peak intensity ($I_0 = 40$~GW/cm$^2$) are shown in Fig.~\ref{fig4}(b). The phase of MD abruptly jumps around the pulse center in such a way so that ED and MD oscillate almost in phase. The corresponding MD and ED amplitudes are shown in the inset in Fig. 4(b). While the ED amplitude smoothly follows the incident pulse shape, the MD amplitude also experiences sharp change at the pulse maximum. At this moment, due to large EHP density the particle is detuned from the MD resonance condition, resulting in abrupt decrease of the MD amplitude.
At this point transition to unidirectional scattering during the rest of the pulse may be expected.

In order to address radiation directivity evolution, we calculate the angular emitted power distribution created by perpendicular electric and magnetic dipoles in E- and H-plane as~\cite{Geffrin2012}
\begin{equation}
{S_E}\left( \theta  \right) = {\left| {\tilde m + \tilde p\cos \theta } \right|^2},~
{S_H}\left( \theta  \right) = {\left| {\tilde p + \tilde m\cos \theta } \right|^2},
\end{equation}
with $\theta=0$ being the forward direction and $\theta=\pi$ backward direction. Fig.~\ref{fig4}(c) shows the Front-to-Back ratio (FBR) defined as ${\text{FBR = }}\tfrac{{{S_E}\left( 0 \right)}} {{{S_E}\left( \pi  \right)}}$ as well as the scattering power patterns for the largest pulse intensity before and after the pulse center arrival.

In the regime of weak intensity ($I_0 \sim 10$~GW/cm$^2$) EHP density is not enough to cause significant refractive index change, so that scattering is dominated by magnetic dipole excitation with low FBR. However, as the pulse intensity increases, EHP creates larger refractive index change. As a result, MD resonance shifts to shorter wavelengths causing significant decrease of the magnetic dipole polarizability $\alpha_m$. At the same time, the electric dipole resonance is barely affected by EHP-induced refractive index change. This results in a Huygens-like behavior with FBR as high as 100 for $40$~GW/cm$^2$ incident pulse, Fig.~\ref{fig4}(c).

Remarkably, switching to unidirectional Huygens regime under irradiation by a strong pulse occurs at 100~fs scale time, Fig.~\ref{fig4}(c). Such fast reconfiguration occurs due to intense EHP generation via two-photon absorption, whose rate scales quadratically with $I_0$. In order to quantitatively describe this feature of transient nanoparticle response, we introduce the switching time $\tau_{\rm switch}$ as the minimal duration of a Gaussian pulse with given peak intensity $I_0$ required to achieve $\text{FBR}=5$ during the pulse action.

This switching time $\tau_{\rm switch}$ is shown in Fig.~\ref{fig5}(a) as a function of peak intensity and excitation wavelength assuming that nanoparticle radius obeys the MD resonance condition, i.e. $2R\approx\lambda \sqrt {  \text{Re} \left( \varepsilon_\text{Si} \right) }$. Clearly, ultrafast reconfiguration with $\tau_{\rm switch} \le 100$~fs can be achieved in the wavelength range $550-650$~nm where high one- and two-photon absorption in silicon enables dense enough EHP generation at the ultrashort time scale (see Supplemantary material for a 2D plot of EHP density). It also shows that the switching time gradually reduces with increasing wavelength in the IR despite very low one- and two-photon absorption of Si in this range. This occurs due to $\lambda^2$ scaling of Drude contribution to the permittivity of photoexcited silicon.

Required intensity and pulse duration depend on the value of desired FBR of a silicon nanoparticle in the switched state. In the regime of exact Huygens source, when $\alpha_e=\alpha_m$, backward scattering cancels completely, but it would require larger fluences. However, finite-FBR although highly directional regimes can be obtained at lower EHP densities and therefore smaller fluence. In Fig.~\ref{fig5}(b) we show the switching intensity assuming constant 200~fs pulse duration for a series of desired FBR values. We also derive a useful analytical estimation of switching intensity for infinitely large FBR (see Supplementary materials):

\begin{equation}
{I_{{\text{switch}}}} = \frac{{{c^2}}}{{4\lambda }}{\left( {\frac{{{\pi ^2}}}{{\operatorname{Re} \left({\varepsilon _{{\text{Si}}}} \right)  }}} \right)^2}\sqrt {\frac{\hbar }{{{e^2}}} \cdot \frac{{\operatorname{Re} \left( {\varepsilon _{{\text{Si}}}} \right)  {m^*}}}{{\tau \operatorname{Im} {\chi ^{(3)}}}}} .
\end{equation}
The estimated value of switching intensity reduces with increasing Im$\chi ^{(3)}$ and $\operatorname{Re}\left( \varepsilon _{\rm Si}\right)$. Therefore, using high-index dielectrics with high third-order nonlinearity is preferable for the optical switching. Particularly, as Fig.~\ref{fig5}(a) suggests, a laser pulse with $\lambda$~=~550--700~nm and $I_0$~$>$~15~GW/cm$^2$ is the most optimal for the ultrafast reconfiguration of the nc-Si nanoantenna. 

In practice, achieving of FBR=$\infty$ is not necessary, whereas FBR=3--5 is sufficient for significant change of the transmitted optical signal through the nanoswitch, yielding relatively low range of incident intensities $I_0$~=~15--60~GW/cm$^2$ at $\lambda$~=~500--1500~nm [see Fig.~\ref{fig5}(b)]. Particularly, at $I_0$~=~15~GW/cm$^2$, $\lambda$~=~550--700~nm, $\tau$~=~100~fs, and realistic focal spot area $\approx$~1~$\mu$m$^2$, 15~pJ pulse is sufficient for considerable switching (FRB$\approx$3) at the ultrafast time. Such energy and intensity levels are far below the damage threshold of Si ($I_{\rm damage}~\sim$~10$^{2}$--10$^{3}$~GW/cm$^2$, Refs.~\cite{Makarov2015,Shcherbakov2015}) for such type of nanoparticles. Interestingly, our simple expression for $I_{\rm switch}$ gives the same $\sim \tau^{-1/2}$ scaling as the damage threshold $I_{\rm damage} \sim \tau^{-1/2}$.\cite{bauerle2013laser} Therefore, the ratio $I_{\rm switch} / I_{\rm damage}$ is approximately constant in a broad range of pulse durations, simplifying the system optimization.

In order to evaluate the ultimate performance of the proposed nanostructure, we note that its bandwith is mainly limited by reverse switching from the photoexcited unidirectional state to dipole-like one during the EHP relaxation. Numerical simulations indicate that such switching from FBR~=~3 to its initial state occurs during approximately 4~ps yielding the maximum bandwidth of about 250~Gbit/s. Our pump-probe measurements confirm the possibility of device operation at this speed.

\section{Conclusion}
To summarize, we have shown that generation and relaxation of EHP in a silicon nanoparticle under non-damage regime of photoexcitation can be as fast as 2.5~ps, being promising for ultrafast all-optical data processing. Basing on the developed model, we have analyzed transient optical dynamics of the nanoparticle, demonstarting accelerated scattering pattern reconfiguration from a dipole-like to a Huygens element-like pattern during laser-nanoantenna interaction. Our results provide a general strategy for the experimental parameters optimization for achieving effective all-optical signal processing by using a single dielectric nanoparticle. (i) The material should have high third order nonlinearity and refractive index at pump wavelength; (ii) Provided that (i) is satisfied, longer wavelengths of the signal are preferable; (iii) The use of nanocrystalline or amorphous state is favorable due to faster relaxation rate as compared to pure crystalline state; (iv) Irradiation in the vicinity of the Kerker condition gives the most dramatic change of the scattering properties. We envision that the proposed approach for ultrafast tuning of Huygens source will be useful for the development of novel ultrafast nonlinear optical nanodevices, where interplay between electric and magnetic modes plays the crucial role in scattering behavior.

\begin{acknowledgments}
We are thankful to Andrey Ionin and Pavel Damilov for assistance with pump-probe measurements and to Yuri Kivshar, Arseniy Kuznetsov, and Maxim Shcherbakov for fruitful discussions. The experimental and theoretical parts of the work were financially supported by grants of Russian Science Foundation �15-19-00172 and 15-19-30023, respectively. D.G.B. acknowledges support from the Russian Foundation for Basic Research (project No 16-32-00444).
\end{acknowledgments}

%

\pagebreak
\newpage
\widetext
\begin{center}
\textbf{\large Supplementary Material: Nonlinear Transient Dynamics of Photoexcited Silicon Nanoantenna for Ultrafast All-Optical Signal Processing}
\end{center}
\setcounter{equation}{0}
\setcounter{figure}{0}
\setcounter{table}{0}
\setcounter{page}{1}
\setcounter{section}{0}
\makeatletter
\renewcommand{\theequation}{S\arabic{equation}}
\renewcommand{\thefigure}{S\arabic{figure}}
\renewcommand{\bibnumfmt}[1]{[S#1]}
\renewcommand{\citenumfont}[1]{S#1}

\section{Calculation of volume-averaged fields}

Calculation of the one- and two-photon absorption rates $W_{1,2}$ requires knowing of the electric field value averaged over the nanoparticle volume, $\left\langle {{{\left| {{\mathbf{\tilde E}}} \right|}^2}} \right\rangle $ and $\left\langle {{{\left| {{\mathbf{\tilde E}}} \right|}^4}} \right\rangle $. We should relate these values to the instantaneous amplitudes of electric and magnetic dipole moments, whose dynamics is governed by Eq.~(1) in the main text. To do that, we assume that at each moment electric field inside the particle can be represented as a sum of electric dipole and magnetic dipole modes:
\begin{equation}
\begin{gathered}
  {\mathbf{E}}({\mathbf{r}}) = {{\mathbf{E}}_{{\text{MD}}}}({\mathbf{r}}) + {{\mathbf{E}}_{{\text{ED}}}}({\mathbf{r}}) = {A_{{\text{MD}}}}{j_1}(\sqrt \varepsilon  kr)\left[ {{{\mathbf{e}}_\theta }\cos \varphi  - {{\mathbf{e}}_\varphi }\sin \varphi \cos \theta } \right] +  \hfill \\
  {A_{{\text{ED}}}}\left[ {2\frac{{{j_1}(\sqrt \varepsilon  kr)}}
{{\sqrt \varepsilon  kr}}{{\mathbf{e}}_r}\cos \varphi \sin \theta  + \frac{{\left( {\sqrt \varepsilon  kr \cdot {j_1}\left( {\sqrt \varepsilon  kr} \right) } \right)'}}
{{\sqrt \varepsilon  kr}}\left( {{{\mathbf{e}}_\theta }\cos \theta \cos \varphi  - {{\mathbf{e}}_\varphi }\sin \varphi } \right)} \right] \hfill \\
\end{gathered}
\end{equation}

The values $A_{\rm MD}$ and $A_{\rm ED}$ are found by integrating the near-field current of the field distribution $E$ and equating the resulting dipole moments to known values of $\tilde p$ and $\tilde m$:
\begin{equation}
{\mathbf{\tilde m}} = \frac{1}
{{2c}}\int_V {{\mathbf{r}} \times \left( { - i\omega } \right)\frac{{\varepsilon  - 1}}
{{4\pi }}{{\mathbf{E}}_{{\text{MD}}}}{d^3}{\mathbf{r}}}
\end{equation}

\begin{equation}
{\mathbf{\tilde p}} = \int_V {\frac{{\varepsilon  - 1}}
{{4\pi }}{{\mathbf{E}}_{{\text{ED}}}}{d^3}{\mathbf{r}}}
\end{equation}
Now, volume-averaged electric field can be directly calculated by integrating expression~(1) over the nanoparticle volume. This procedure is repeated at each step of numerical simulations.

\section{Permittivity of excited silicon}

This is a well-know result and can be found, e.g., in Refs.~\cite{Sokolowski-Tinten2000S, Makarov2015S}:
\begin{equation}
\varepsilon(\omega,\rho_{\rm eh})=\varepsilon_{\rm IB}(\omega^*)\left(1-\frac{\rho_{\rm eh}}{\rho_{\rm bf}}\right)
-\frac{\omega^2_{\rm pl}(\rho_{\rm eh})}{\omega^2+1/(\tau^2_e(\rho_{\rm eh}))}\left(1-\frac{i}{\omega\tau_e(\rho_{eh})}\right),
\label{eqEps}
\end{equation}
where the above mentioned $\rho_{\rm eh}$-dependent bandgap shrinkage effect on interband transitions is accounted by introducing effective photon frequency $\omega^{*}= \omega+\Theta\rho_{\rm eh}/\rho_{\rm bgr}$. Here, the characteristic renormalization EHP density $\rho_{\rm bgr}\approx1\times10^{22}$~cm$^{-3}$, the factor $\Theta$ is typically about 5$\%$ of the total valence electron density ($\approx2\times$10$^{23}$~cm$^{-3}$ in Si) to provide the ultimate 50$\%$
electronic direct bandgap renormalization, i.e., $\hbar\Theta\approx$ 1.7~eV of the effective minimal gap $\approx3.4$~eV in silicon, and $\rho_{\rm bf}$ is the characteristic band capacity of the specific photo-excited regions of the first Brillouine zone in the k-space (e.g., $\rho_{\rm bf}(L)\approx4\times10^{21}$~cm$^{-3}$ for L-valleys and $\rho_{\rm bf}(X)\approx 4.5\times10^{22}$ cm$^{-3}$ for X-valleys in Si), affecting interband transitions via the band-filling effect~\cite{Sokolowski-Tinten2000S}. The bulk EHP frequency $\omega_{\rm pl}$ is defined as
\begin{equation}\label{S4}
\omega_{\rm pl}^{2}(\rho_{\rm eh})=\frac{\rho_{\rm eh}e^2}{\varepsilon_0\varepsilon_{\rm hf}(\rho_{\rm eh})m^{*}_{\rm opt}(\rho_{\rm eh})},
\end{equation}
where the averaged over L- and X- valleys effective optical (e-h pair) mass $m^{*}_{\rm opt}\approx0.18$$m_{\rm e}$ (Ref.~\cite{Sokolowski-Tinten2000S}) is a $\rho_{\rm eh}$-dependent quantity, varying versus transient band filling due to the band dispersion and versus bandgap renormalization. The high-frequency electronic dielectric constant $\varepsilon_{\rm hf}$ was modeled in the form $\varepsilon_{\rm hf}(\rho_{\rm eh})=1+\varepsilon_{\rm hf}(0)\times\exp(-\rho_{\rm eh}/\rho_{\rm scr})$, where the screening density $\rho_{\rm scr}\approx1\times10^{21}$~cm$^{-3}$ was chosen to provide $\varepsilon_{\rm hf}\rightarrow1$ in dense EHP. The electronic damping time $\tau_e$ in the regime of dense EHP at the probe frequency $\omega_{\rm pr}$ was taken, similarly to metals, in the random phase approximation as proportional to the inverse bulk EHP frequency $\omega_{\rm pl}^{-1}$. Here, $\tau_e$ is evaluated for $\hbar\omega>k_{\rm B}T_{\rm e}$ in the form $\tau_{\rm e}(\rho_{\rm eh})\approx3\times10^{2}/(\omega_{\rm pl}(\rho_{\rm eh}))$, accounting multiple carrier scattering paths for the three top valence sub-bands, and multiple X-valleys in the lowest conduction band of silicon.


\begin{figure}
\includegraphics[width=.9\columnwidth]{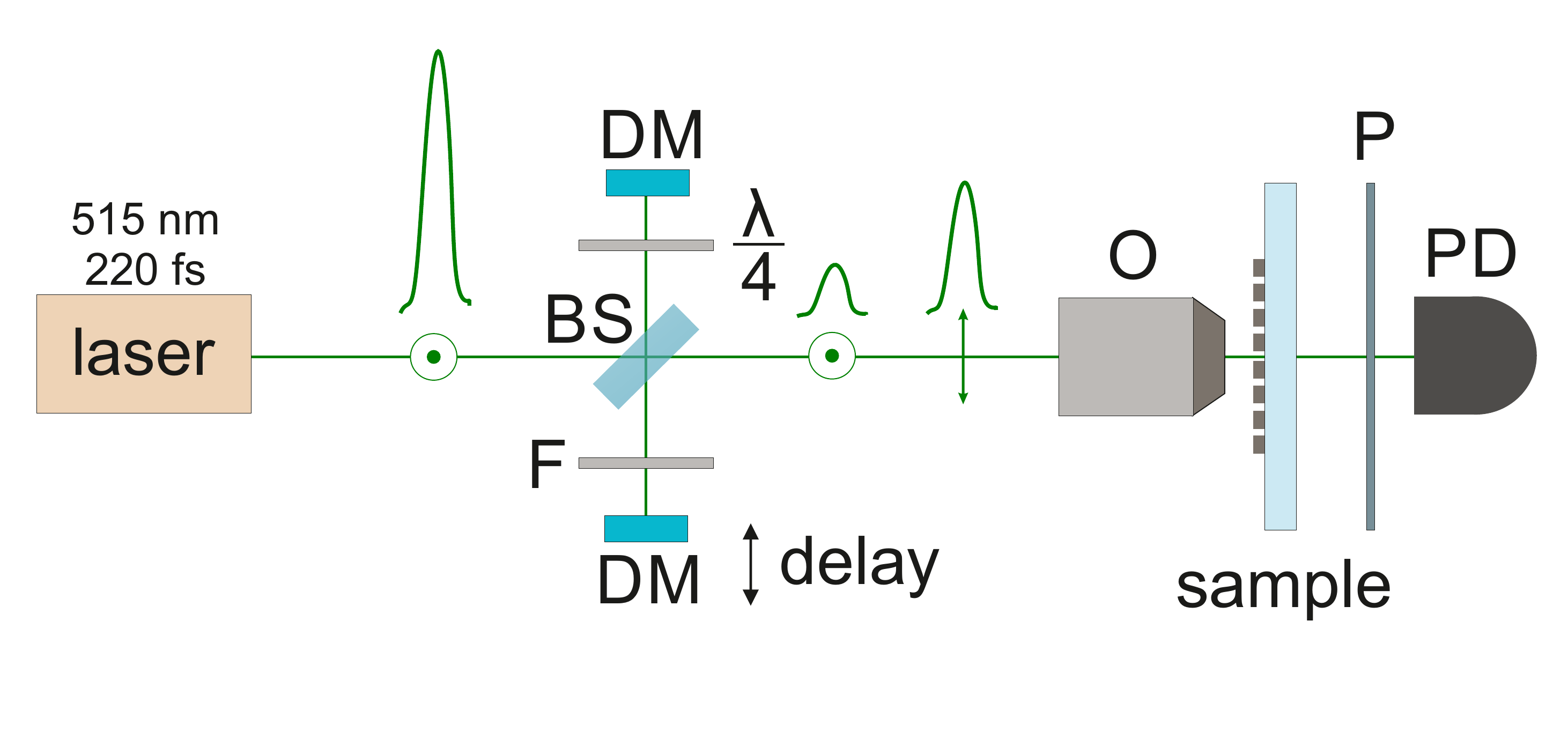}
\caption{Experimental setup for the pump-probe transmittance measurements, where BS is beam splitter, DM is dielectric mirror, O - objective, P - polarizer, PD - photodetector.}
\label{figS1}
\end{figure}

\section{Optical characterisation}

The broadband spectral measurements of the reflected signal from a single nanoparticle ($R_{s}$) were carried out by means of strong focusing and collection of light ($\lambda$=400--900 nm) to a spectrometer (Horiba LabRam HR) through an achromatic objective with a numerical aperture NA=0.95.

Raman scattering data were measured by a micro-Raman apparatus (Raman spectrometer HORIBA LabRam HR, AIST SmartSPM system). As a source 632.8-nm HeNe laser were used. The Raman spectra were recorded through the 100$\times$ microscope objective (NA=0.9) and projected onto a thermoelectrically cooled charge-coupled device (CCD, Andor DU 420A-OE 325) with a 600-g/mm diffraction grating. Individual nanoparticle spectrum was recorded by a commercial spectrometer (Horiba LabRam HR) when the nanoparticle was precisely places (accuracy about 100 nm) in the center of the laser beam (0.86-$\mu$m diameter) focused on the substrate.

\section{Time-domain transmittance measurements}

We exploit a laser pulses at 515~nm wavelength and 1~kHz repetition rate, which are split by two pulses with orthogonal linear polarizations and have 10-times difference between their intensities. Maximum energy of the pump pulse in non-damage regime is about 10~nJ. The pulses are focused by a NA=0.25 objective onto the nanoparticles array, providing focused laser spot size of about $r_{1/e}\approx$~2.5~$\mu m$. The energy of the transmitted probe pulse is measured by a GaP photodetector, whereas the pump pulse is filtered by a polarizer.

\begin{figure}
\includegraphics[width=.7\columnwidth]{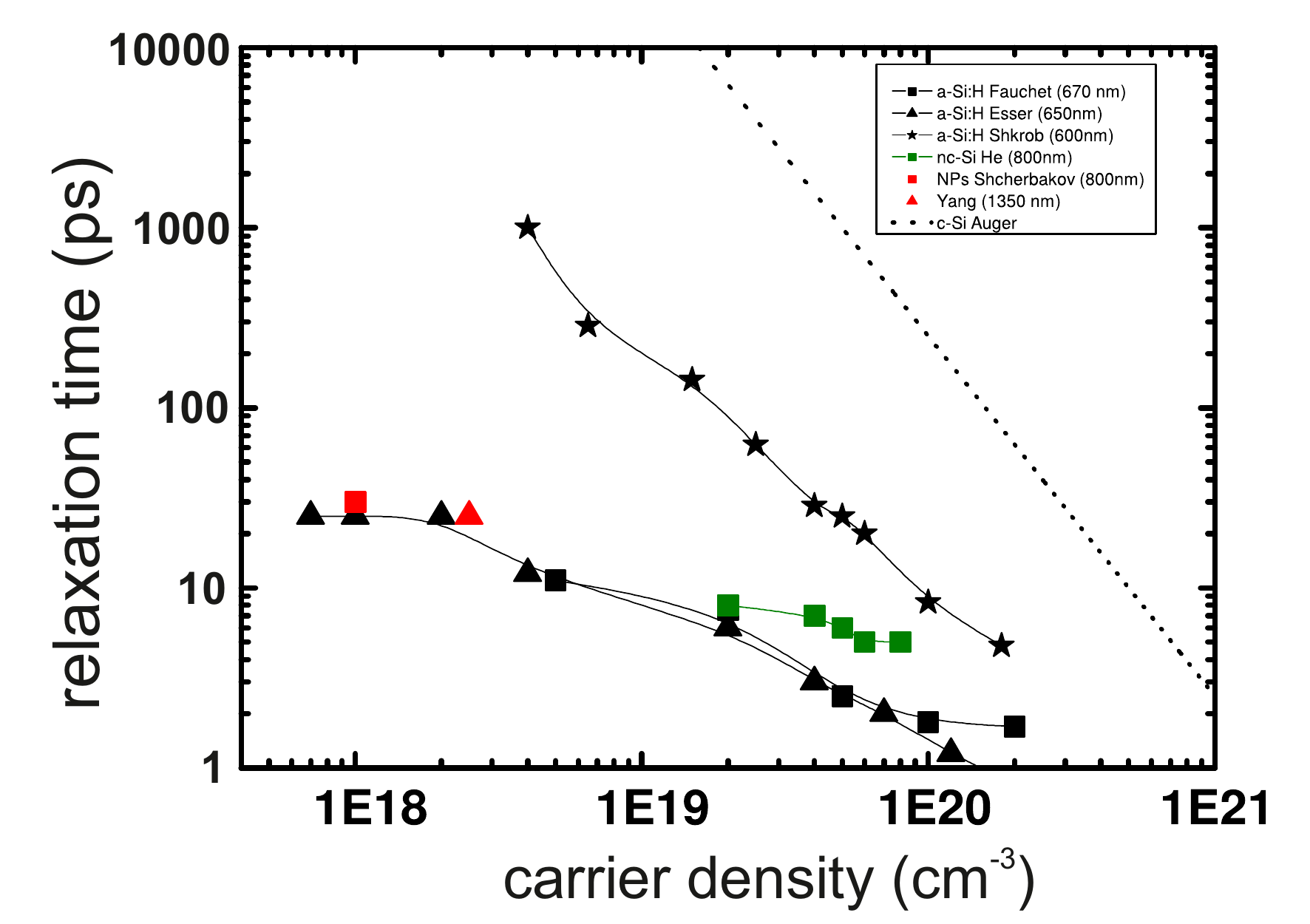}
\caption{Relaxation time of EHP as a function of EHP density. Data from~\cite{Shcherbakov2015S} (red square),~\cite{yang2015nonlinearS} (red triangle),~\cite{Optlett15S} (green squares),~\cite{Shkrob1998S} (black stars),~\cite{Esser1993S} (black triangles),~\cite{Fauchet1992S} (black squares), and calculated Auger relaxation for c-Si (black dotted line).}
\label{figS3}
\end{figure}

\section{Recombination}

For silicon, the carrier trapping ($\Gamma _{\rm TR}$) on point defects~\cite{JepsenS} and self-trapping~\cite{MaoS} is well known to be the dominating mechanism at low EHP densities ($<10^{20}$~cm$^{-3}$) and low crystallinity. So-called bi-molecular ($\Gamma _{\rm BM}$) non-radiative and radiative recombination (both resonant and non-resonant) processes govern relaxation dynamics mostly in amorphous silicon~\cite{esserS, fauchetS}. For crystalline silicon in a broad range of EHP densities, the relaxation proceeds via Auger recombination ($\Gamma_{\rm A}$)~\cite{ShankS}. However, different methods of silicon fabrication yield the materials with different defects types and concentrations, resulting in different coefficients for ${\Gamma_{\rm TR}}$, ${\Gamma _{\rm BM}}$, and ${\Gamma _{\rm A}}$~\cite{esserS, fauchetS, Optlett15S}. Fig.~\ref{figS3} represents some experimental data from previous works on EHP relaxation in different silicon-based materials and nanostructures~\cite{Fauchet1992S, Esser1993S, Shkrob1998S, Optlett15S, yang2015nonlinearS, Shcherbakov2015S} and their comparison with Auger recombination in c-Si given as $\tau=(4\cdot$~10$^{ - 31}$~$\rho_{\rm eh}^2$)$^{-1}$~s, revealing much faster relaxation dynamics for amorphous and nanocrystalline states as compared to pure crystalline silicon.

\begin{figure}[!t]
\includegraphics[width=.7\columnwidth]{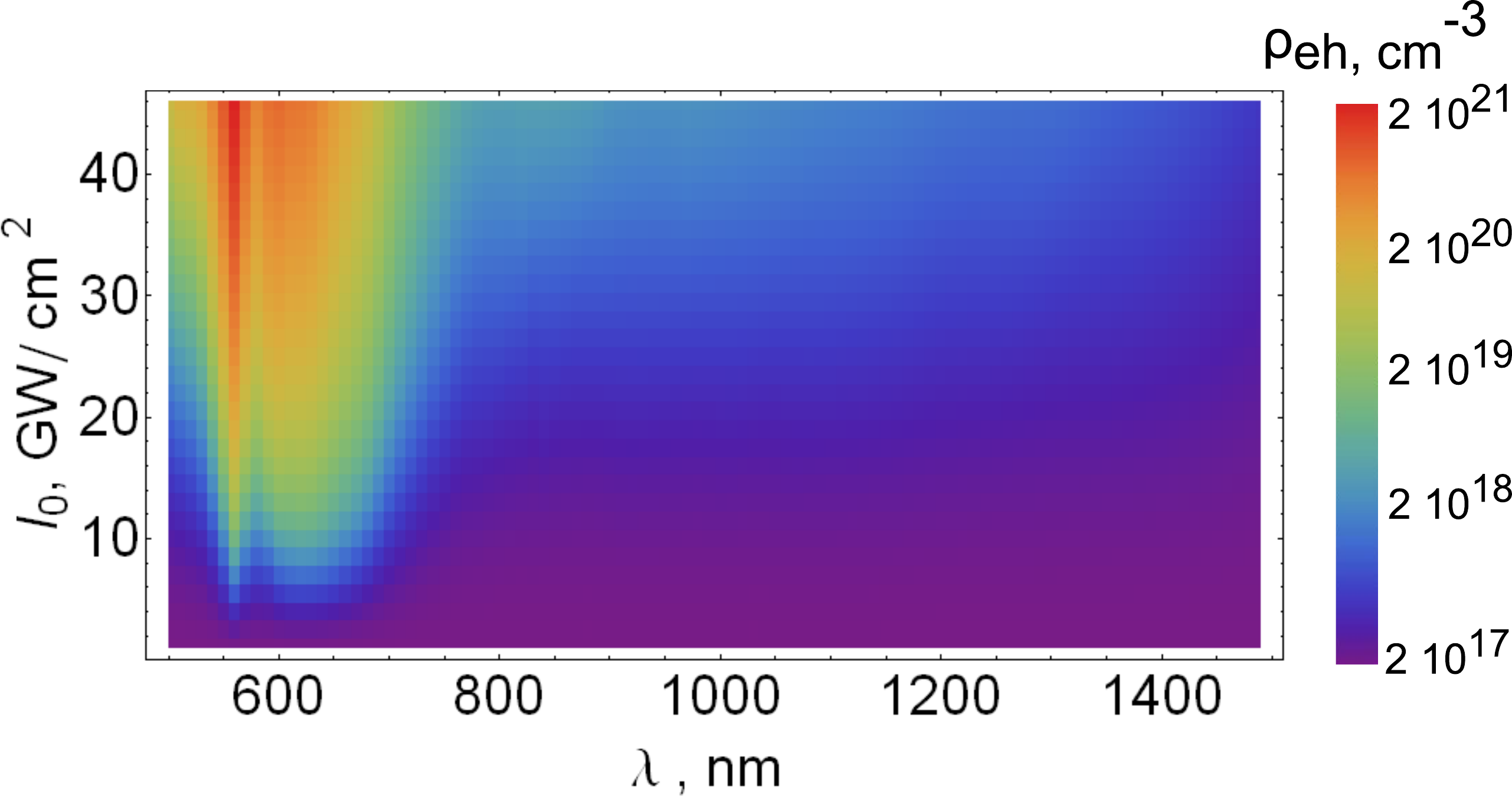}
\caption{Highest value of EHP density inside a resonant nanoparticle during the pulse action as a function of radius and incident intensity.}
\label{figS4}
\end{figure}
\section{Generated EHP density}
In Fig.~\ref{figS4} we show the highest value of EHP density attainable during the pulse action assuming that nanoparticle radius obeys MD resonance condition.

\section{Estimation of switching intensity}
Developed theory allows us to obtain an evaluation of the threshold pulse intensity $I_{\rm switch}$ necessary for transition to the unidirectional scattering regime. Assume that pulse wavelength $\lambda$ is tuned to the magnetic dipole resonance of unexcited particle. The first Kerker condition, when $\alpha_e=\alpha_m$, is satisfied at ${\lambda _{{\text{Kerker}}}} \approx 2.2R\sqrt {{\varepsilon_0}} $~\cite{Lukyanchuk2015S}. Given the constant excitation wavelength, we conclude that refractive index of the photoexcited silicon should be decreased by 10$\%$ to achieve the Kerker's condition: $\delta n \approx -0.1n,~  \delta \varepsilon  = \delta \left( {{n^2}} \right) \approx - 0.2\varepsilon_0$.
The main contribution to modified permittivity at optical and IR frequencies is provided by the free carriers term in Eq.~(4) and it can be roughly estimated as $\delta \varepsilon  \approx \delta( - {{\omega _{\rm pl}^2} \mathord{\left/  {\vphantom {{\omega _{\rm pl}^2} {{\omega ^2}}}} \right.  \kern-\nulldelimiterspace} {{\omega ^2}}})$, where $\omega_{\rm pl}$ is the bulk EHP frequency defined as
$\omega _{{\text{pl}}}^2 = 4\pi {\rho _0}{e^2}/\left( {{\varepsilon _{{\text{hf}}}}m_{{\text{opt}}}^*} \right)$.
Here, the effective optical (e-h pair) mass $m^{*}_{\rm opt}\approx0.18$$m_{\rm e}$ and high-frequency electronic dielectric constant $\varepsilon_{\rm hf}\rightarrow1$ in dense EHP~\cite{Makarov2015}.

To estimate peak EHP density $\rho_0$ created by the incident pulse, we note that during a typical $100-200$~fs pulse duration EHP relaxation can be neglected as it occurs on a ps scale. Further, we notice that two-photon absorption is more efficient at characteristic intensities required for activation of Huygens regime. Therefore we integrate the EHP rate equation (Eq.~(2) of the main text) and estimate peak EHP density as ${\rho _0} \approx \frac{\tau }{{16\pi \hbar }}\operatorname{Im} {\chi ^{(3)}}{\left| {{E_{{\text{in}}}}} \right|^4}$.
Finally, recalling that at magnetic dipole resonance ${\alpha _m} = \tfrac{{3i}}{{2{k^3}}}$, and that electric field inside the particle is enhanced by a factor of $F \sim 6{\alpha _m}/(k{R^4}\varepsilon )$, we substitute this formula into expression for $\delta \varepsilon$ and arrive at the desired evaluation of threshold intensity:
\begin{equation}
{I_{{\text{switch}}}} = \frac{{{c^2}}}
{{4\lambda }}{\left( {\frac{{{\pi ^2}}}
{{\operatorname{Re} ({\varepsilon _{{\text{Si}}}})}}} \right)^2}\sqrt {\frac{\hbar }
{{{e^2}}} \cdot \frac{{\operatorname{Re} ({\varepsilon _{{\text{Si}}}}){m^*}}}
{{\tau \operatorname{Im} {\chi ^{(3)}}}}} 
\label{eqEval}
\end{equation}

\end{document}